\documentclass{acmconf}
\usepackage{multicol}
\usepackage{epsfig}
\usepackage{amsmath, amsthm, amssymb}

\begin{document}
\title{A Game Theoretical Approach to Modeling Full-Duplex Information Dissemination}
\author{%
Dmitry Zinoviev and Vy Duong\\%
Mathematics and Computer Science Department, Suffolk University\\%
Boston, Massachusetts 02114, USA%
}
\date{}
\maketitle


{\bf Keywords}: Social Network, Game Theory, Learning, Knowledge, Popularity, Reputation, Rumor.
\vskip\baselineskip

\abstract{
One major function of social networks (e.g., massive online social networks) is the dissemination of information such as scientific knowledge, news, and rumors. Information can be propagated by the users of the network via natural connections in written, oral or electronic form. The information passing from a sender to a receiver intrinsically involves both of them considering their self-perceived knowledge, reputation, and popularity, which further determine their decisions of whether or not to forward the information and whether or not to provide feedback. To understand such human aspects of the information dissemination, we propose a game theoretical model of the two-way full duplex information forwarding and feedback mechanisms in a social network that take into account the personalities of the communicating actors (including their perceived knowledgeability, reputation, and desire for popularity) and the global characteristics of the network. The model demonstrates how the emergence of social networks can be explained in terms of maximizing game theoretical utility.
}

\section{INTRODUCTION}
A social network is an ensemble of communicating personalities based on the concept of social proximity.  The members of a social network can form communities~\cite{spertus2005}, influence other members~\cite{crandall2008}, and engage in other social activities. 
One major function of social networks (in particular, massive online social networks) is the dissemination of information such as scientific knowledge, news, and rumors~\cite{1099206,nekovee2008,watts2002,zanette2002}. As an important form of social organization, information can shape public opinion, inform and misinform the society, cause panic in a society, promote products, etc.~\cite{nekovee2008}.
Information can be propagated by the members of the network via natural connections in written, oral or electronic form. 

Due to its importance, information dissemination or diffusion has been one of the focuses in social network research. For example, theories of rumor spreading are proposed in ~\cite{nekovee2008, zanette2002} to study the information dissemination.
Game theoretical approach to information propagation (namely, to learning) has been suggested by Gale {\it et al.}, Acemoglu {\it et al.}~\cite{NBERw14040,gale2003}. Ellwardt and van Duijn explored gossiping in small (organizational) social networks~\cite{ellwardt2009}.
Since information dissemination (and other various social network activities)
are supported by the structural organization of social networks, social network topology receives a lot of research attention. For example, 
an effect of network topology on the information diffusion was observed in~\cite{hirshman2009}: sparse networks are more effective for information entrance, and clustered (cellular) network structure decreases information diffusion.

In this paper, we propose a game theoretical model for the information passing between two members of a social network. The novelty of our model is that psychological characteristics are explicitly modeled in information dissemination. 
In our model, information passing intrinsically involves both parties considering their psychological characteristics: self-perceived knowledge, reputation, and popularity---which further determine their decisions of whether or not to forward the information and whether or not to provide feedback. The decisions are also based on the global properties of the network, such as the overall quality of information and the way unreliable rumors are treated by the network members.

Feedback in information dissemination is explicitly considered as strategic moves in our game theoretical model. Related to our work, Lampe {\it et al.} also analyzed the mechanism of feedback, its influence on the members of online communities, and its role in learning transfer~\cite{1099206}. Similar concept of social influence, but in the context of community building, has been researched by Crandall {\it et al.}~\cite{crandall2008}.

This paper is an extension of the results described in ~\cite{zinoviev2010}. In the original paper, we presented an analysis of a one-directional atomic communication, where one actor is a speaker (he sends a message) and the other is a listener (and optionally a commenter---he receives the message and sends a reply), and the comment, if any, is indivisible from the original message. However, in real life people can engage in a concurrent conversation: e.g., Alice may post a note on Bob's Facebook wall, and Bob may post a note on Alice's wall at the same time, without knowing yet about her post. This interaction cannot be represented as a sequence of atomic one-way interactions. The current version of our model allows two actors to communicate in a full duplex mode, when either or both actors are speaking, listening, and commenting at the same time.

The model that we developed can be used to explain communication patterns in massive online social networks (MOSN) and even the formation of a MOSN based on the principle of the highest utility.

To the best of our knowledge, no existing paper takes a similar approach.

\section{\label{model}INFORMATION DISSEMINATION MODEL}

In this section, we present our model for characterizing actors and for the information dissemination between two actors. Each actor has a utility function which is a combination of her knowledge, reputation, and popularity. The information passing between a pair of actors involves learning, feedback, and utility updating of both participants. We propose a game theoretic model to study the strategic actions of actors in Section~\ref{game_model} and then further look at the information dissemination in a network setting involving a large number of actors and at a possibility of social network emergence in Section~\ref{numerical}.

\subsection*{Assertions, Beliefs, and Knowledge}

We suppose that there is a discrete finite set $\boldsymbol{F}$ of $N$ assertions that is to be shared among all actors. We assume that all assertions are equally important.

An assertion intrinsically can be true or false. We use $\varphi$ to denote the probability that a randomly selected assertion from the set $\boldsymbol{F}$ is true. Note that $\varphi$ is a system-wide parameter. The value of $\varphi$ is higher for formal communities (e.g., a scientific community) and lower for informal communities (e.g., a chatroom).

In addition to its intrinsic objective truthfulness, each assertion known to an actor has a subjective belief associated with it: the actor either believes that the assertion is true or false, or fails to make a decision for herself. An assertion that an actor does not know to be true or false is essentially a rumor---a ``story\ldots without any known authority for [its] truth''~\cite{webster1913}.

An actor $A_i$ knows $F_i\le N$ assertions, among which $A_i$ believes $F_i^+$ assertions are true and $F_i^-$ assertions are false, and treats all other known assertions $F_i^\circ$ as rumors (obviously, $F_i^++F_i^\circ+F_i^-=F_i$). The probability of a randomly chosen assertion $\Phi\in\boldsymbol{F}$ (from the whole assertion set for the network) being known by $A_i$ is given by $f_i=F_i/N$. Similarly, we can find the probability of an assertion believed as true, rumor, or false as $f_i^+=F_i^+/F_i$, $ f_i^\circ=F_i^\circ/F_i$, and $f_i^-= F_i^-/F_i$, respectively.

Based on its own known set of assertions $\boldsymbol{F}_i\subseteq\boldsymbol{F}$, actor $A_i$ forms its self-perceived \emph{knowledge} $K_i$ as follows:
\begin{equation}
 K_i=F_i^++F_i^-+\lambda F_i^\circ
\end{equation}
Here, $0\le\lambda\le1$ is a ``rumor discount'' coefficient to capture the extent that an actor is willing to treat rumors as part of her knowledge. An actor with $\lambda=0$ completely discards rumors from her self-perceived knowledge, whereas an actor with $\lambda=1$ treats all her known rumors as if they were trustworthy assertions. Similar to $F_i$, knowledge $K_i$ is also bounded by $N$, and $F_i\le K_i\le N$.
In this paper, we assume that the value of $\lambda$ is the same for all actors in a social network.
We further normalize an actor's knowledge as $k_i=K_i/N$.

It is important to understand that the truthfulness of an assertion is not necessarily in agreement with the actor's beliefs. Good examples of the disagreement would be so-called ``urban myths'' that are intrinsically false, but perceived as true by many people.

Based on their perceived knowledge, we roughly classify all actors into ``Ignoramuses'' (low $k_i$), ``Mediocres'' (medium-ranged $k_i$), and ``Gurus'' (high $k_i$). No actor in the network can definitely tell whether an assertion is true or false. However, we assume that there exists an external verification mechanism (an ``oracle'') that knows the definite answer.

\subsection*{Utility and Personality}

A single communication between any two actors consists of an exchange of at most two assertions (from Alice to Bob and back) and optional feedback message(s).  The communication intrinsically involves both actors considering their \emph{self-perceived knowledge, reputation, and popularity}, which further determine their decisions of whether or not to forward the assertions in the first place and whether or not to provide feedback. An actor's self-perceived knowledge, reputation, and popularity collectively form her \emph{utility}. The weights an actor puts on these three utility components characterize this actor's \emph{personality}. We will give a detailed explanation of these concepts in this section.

We use a non-negative real number to measure an actor $A_i$'s \emph{reputation} $R_i$---``overall quality or character as seen or judged by people in general''~\cite{webster1913}.
Lower values of $R_i$ mean lower trustworthiness in $A_i$, suggesting that opinions expressed by $A_i$---such as information and feedback---were questionable in the past and should be considered with a grain of salt. In the extreme case of $R_i=0$, actor $A_i$'s beliefs are completely not credible. On the contrary, higher values of $R_i$ mean that $A_i$'s subjective evaluation of assertions has been regarded as historically highly credible.

We measure an actor's \emph{popularity} $P_i$ using another non-negative real number. An actor's popularity in a society is somewhat synonymous to social visibility~\cite{parkhurst98}: $P_i=0$ corresponds to an actor who does not speak in public
and in general is not even known to exist. An actor with high popularity enjoys popular attention. We assume that there is no correlation between $P_i$ and $R_i$ for the same actor $A_i$. We suppose that an actor's popularity decays by $\Delta P=-\delta$  per unit time unless the actor participates in information exchange with other actors.

We further define normalized reputation and normalized popularity as $r_i=R_i/N$ and $p_i=P_i/N$. Note, in our study, we choose $N$ to be sufficiently large such that both $r_i$ and $p_i$ are no larger than one.

We believe that the purpose of a rational actor is to maximize her utility $U_i$, defined as a convex combination of knowledge, reputation, and popularity with coefficients $0\le\kappa,\rho,\pi\le1,\kappa+\rho+\pi=1$: 
\begin{equation}
 U_i=\kappa K_i+\rho R_i+\pi P_i.\label{payoff}
\end{equation}

We use a particular set of coefficients $\{\kappa,\rho,\pi\}$ to characterize a particular type of actors' personality. For example, $\kappa=\rho=0,\pi=1$ describe a network of ``Internet trolls'' (actors, for whom bloated popularity is the primary goal of networking). On the other hand, $\kappa=\rho=0.5,\pi=0$ probably corresponds to a scientific community of knowledge seeking altruists who care about their reputation and wisdom, but not about being quoted or even published.

In this paper, we focus on a homogeneous network where all actors' have the same utility function. We understand that in a real social network, actors are heterogeneous. We leave the heterogeneous network as future work.

\subsection*{Information Transmission between a Pair of Actors}

During the passing of an assertion between two actors, both actors update their self-perceived knowledge, reputation, and popularity. This process involves evaluating knowledge, learning assertions, sending feedback, and updating reputation and popularity. We now describe in this section the basic steps that are performed in this process, but leave the discussion on the actors' the strategic decision making in the next section.

We assume that initially each actor pre-learns a random collection of assertions, which she randomly classifies as true assertions, false assertions or rumors, and that a communication (including both sending assertions and feedback) always takes a unit time.

\subsubsection*{Evaluating Knowledge}

When an actor $A_i$ has an assertion $\Phi\in \boldsymbol{F}_i$ to share with her neighbor, she will decide whether to forward $\Phi$ to a neighbor or hold it in order to maximize her utility, as defined by Eq.~(\ref{payoff}). If $A_i$ sends $\Phi$ to her neighbor $A_j$, then $A_j$ may choose to respond to $A_i$ with either positive or negative feedback $\Psi$. Simultaneously, $A_j$ has to make a similar decision on whether to forward his assertion to $A_i$.

Upon receiving $\Phi$ from the $A_i$, the other actor $A_j$ may or may not be able to decide if $\Phi$ is true or false, based on the following considerations:
\begin{itemize}
\item $A_j$'s self-perceived knowledge.
\item The probability of $\Phi$ being true by nature (an intrinsic or objective characteristic but not a subjective belief by any actor), given by the system-wide parameter $\varphi$.
\item $A_i$'s reputation, $R_i$.
\item $A_i$'s opinion on $\Phi$: for an arbitrary assertion, the probability of that assertion being perceived by $A_i$ as true or false is given by $f_i^+$ or $f_i^-$; $A_i$ has no definite opinion on the assertion with the probability of $f_i^\circ$.
\end{itemize}

Let $g^+$ and $g^-$ be the probabilities of $A_j$ deciding that $\Phi$ is true or false respectively, and $g^\circ$ be the probability of $A_j$ failing to decide on $\Phi$ (i.e. declare it as a rumor). Note that $g^++g^-+g^\circ=1$, because eventually $A_j$ has to make some decision. If both $A_i$ and $A_j$ have complete knowledge of all assertions (i.e., $k_i=k_j=1$), then we have:
\begin{equation}
\begin{split}
g^+&=f_i^+=f_j^+=\varphi,\\
g^-&=f_i^-=f_j^-=1-\varphi,\\
g^\circ&=f_i^\circ=f_j^\circ=0.\label{ideal}
\end{split}
\end{equation}
If $A_j$ is an Ignoramus ($k_j=0$), we assume that $A_j$ chooses to trust $A_i$'s opinion $f_i^*$ ($*\in\{+,-,\circ\}$), discounted by her reputation $r_i$. That is,
\begin{equation}
\begin{split}
g^+&=r_i f_i^+,\\
g^-&=r_i f_i^-,\\
g^\circ&= 1-g^+-g^-.
\end{split}
\end{equation}
In all other cases where ($0<k_j<1$), we assume that  $g^\pm$ are weighted sums defined as:
\begin{equation}
\begin{split}
g^+&=k_j\varphi+\left(1-k_j\right)r_i f_i^+,\\
g^\circ&=\left(1-k_j\right)\left(r_i f_i^\circ+\left(1-r_i\right)\right),\\
g^-&=k_j\left(1-\varphi\right)+\left(1-k_j\right)r_i f_i^-,\label{g-def}
\end{split}
\end{equation}
where $A_j$'s knowledge $k_j$ is a weighting factor.

Note, in the above probability computation, we can guarantee that $g^+,g^-, g^\circ$ are no larger than 1 as $r_i$ is no larger than 1 (as mentioned before, we choose $N$ to be sufficiently large).

\subsubsection*{Learning Assertions and Updating Knowledge and Popularity}

Once an assertion is transmitted from $A_i$ to $A_j$ (or back), both actors' knowledge and popularity may change due to the transmission: $A_i$ informs $A_j$ of a potentially new assertion $\Phi$, and $A_j$ corrects $A_i$'s opinion on $\Phi$. When $A_j$ receives an assertion $\Phi$, one of the following three scenarios can happen:
\begin{enumerate}
 \item $A_j$ knows about $\Phi$, and his new opinion on the assertion, $g^*$, matches his existing opinion $f_j^*$. In this case, $A_j$ is not interested in $\Phi$. He discards the assertion and does not improve $A_i$'s popularity. The probability of this scenario is $f_j\left(g^-f_j^-+g^+f_j^++g^\circ f_j^\circ\right)$.
\item $A_j$ knows about $\Phi$, but will reconsider his belief about $\Phi$. In this scenario, $A_j$ re-labels $\Phi$ in the assertion set $\boldsymbol{F}_j$ and gives $A_i$ a popularity increase of $1$. Re-labeling $\Phi$ does not change $F_j$, but it can change $K_j$ (when $\Phi$ becomes a rumor or ceases to be a rumor). The probability of this event is $f_j\left(1-\left(g^-f_j^-+g^+f_j^++g^\circ f_j^\circ\right)\right)$.
\item Finally, $\Phi$ can be completely new to $A_j$. Then $A_j$ stores $\Phi$ in the assertion set $\boldsymbol{F}_j$ and gives $A_i$ a popularity increase of $1$. The number of assertions of $A_j$'s increases, and the amount of knowledge of $A_j$'s increases, too. This scenario happens with  probability $1-f_j$.
\end{enumerate}
Then, we see that the overall knowledge change at $A_j$'s side is given by:
\begin{equation}
\begin{split}
 \Delta K_j=&\lambda\left(1-f_j\right)\\
+&\left(1-\lambda\right)\left(g^-+g^+-f_j\left(f_j^++f_j^-\right)\right).\label{delta-KR}
\end{split}
\end{equation}

Notice that if $A_i$ knows {\em a priori} that $A_j$ is already in the possession of the assertion that she is about to share with him, then the probability of the above third scenario is $0$, and~(\ref{delta-KR}) changes accordingly:
\begin{equation}
\begin{split}
 \Delta K_j'=\left(1-\lambda\right)\left(g^-+g^+-f_j\left(f_j^++f_j^-\right)\right).\label{delta-KR1}
\end{split}
\end{equation}
This situation may arise when $A_i$ has reinterpreted the assertion. We will not use~(\ref{delta-KR1}) in our further analysis.

Unlike  $A_j$, $A_i$ updates her knowledge based on the feedback provided by $A_j$ (if any). The number of assertions at her side does not change, only $A_i$'s opinion on them can change, as well as $A_i$'s perceived knowledge that depends on the opinion. If $A_i$ trusts $A_j$ (because of $A_j$'s high reputation $R_j$), then $A_i$ can change her belief about $\Phi$. The change is given by the formula:
\begin{equation}
\begin{split}
 \Delta K_i=r_j\left(1-\lambda\right)\left(g^-+g^+-\left(f_i^++f_i^-\right)\right).
\end{split}
\end{equation}

The total popularity premium to $A_i$ is given by:
\begin{equation}
 \Delta P_i=1-f_j\left(g^+f_j^++g^-f_j^-+g^\circ f_j^\circ\right).\label{delta-P}
\end{equation}
$A_j$ enjoys the popularity growth of $1$ only if $A_j$ chooses to send feedback to $A_i$ (i.e., by commenting her original assertion).

\subsubsection*{Sending Feedback and Updating Reputation}

A feedback mechanism is used by $A_j$ to affect the reputation of $A_i$ and, eventually, $A_j$'s own reputation.

We use an actor $A_i$'s reputation $R_i$ as the measure of $A_i$'s ability to inspire belief. Actor $A_i$'s successful prediction of the true nature of an assertion should increase $R_i$, while an incorrect prediction should decrease it. Unfortunately, no actor in the network knows the true value of a random assertion (even a Guru can predict that an assertion is true only with the probability of $\varphi$). That is why we need an external oracle that compares $A_j$'s perception of assertion $\Phi$ with the true nature of that assertion and concludes if the evaluation was successful or not. If $A_j$ is a Guru (high $k_j$), he makes the right choice and earns a reputation increase of $1$. If $A_j$ is an Ignoramus, the best he can do is to trust $A_i$ (to the extent of her reputation). In the latter case, $A_j$ gets a reputation boost of $\left(1\times r_i\right)$ if $A_i$'s belief of $\Phi$ is accurate (which happens with the probability of $\varphi f_i^++(1-\varphi)f_i^-$), and a penalty of $\left(-1\times r_i\right)$ otherwise (with the probability of $\varphi f_i^-+(1-\varphi)f_i^+$).  After an assertion is passed between the actors, the expected change of $A_j$'s reputation is:
\begin{equation}
\Delta R_j=k_j+\left(1-k_j\right)\left(r_i\left(2\varphi-1\right)\left(f_i^+-f_i^-\right)\right).
\end{equation}

In the mean time, $A_j$ can affect $A_i$'s reputation by providing feedback in the following way: if both actors agree on their perceived nature of $\Phi$ (the true nature of $\Phi$ is not involved) and $A_j$ is a credible authority himself, then $A_i$'s reputation improves by value $1$; otherwise, it decreases by value $1$. In other words:
\begin{equation}
\begin{split}
\Delta R_i=r_j\Big(&\left(1-2g^+-2g^-\right)\left(1-2f_i^+-2f_i^-\right)\\
&-2\left(f_i^+g^++f_i^-g^-\right)\Big).\label{delta-CS}
\end{split}
\end{equation}

It is quite possible that the change of knowledge $K$, numbers of known, true, false, and neutral assertions $F$, $F^+$, $F^-$, $F^\circ$, popularity $P$ or reputation $R$, resulting from a communication, will make one or more of these values greater than $N$ or less than 0. Our model is not designed to handle these situations and should be used only when each of these numbers is greater than 1 and less than $N-1$. This anomaly can be avoided by choosing $N$ to be large.

\section{TWO-PLAYER INFORMATION DISSEMINATION GAME}\label{game_model}

In the previous section, we gave a detailed analysis of the basic steps involved in the information transmission between two actors and described how the two actors update their utilities (including self-perceived knowledge, reputation, and popularity) depending on whether or not assertions and feedback messages are transmitted. However, we have not answered this question yet: under what circumstances are the actors willing to transmit assertions and send feedback? In this section, we will address this question under the assumption that both actors know that each of them attempts to maximize its own utility, and they are fully aware of the impact on their own utilities from any combination of their individual choices. 

Such a strategic interaction between the actors can be naturally modeled as a game with both actors being players.
More specifically, both actors play a non-cooperative non-zero-sum rectangular game. The utility change of both players in the game are given by the payoff matrix $M_{mn}, 0\le{m,n}\le4$, and depends on the strategies $S^m_i$ and $S^n_j$ selected by the players. The rows of the matrix correspond to the available actions of $A_i$, and the columns correspond to the available actions of $A_j$. Each actor has to choose between four actions:
\begin{enumerate}
\item[$S^0$:] not to send an assertion and not to provide feedback,
\item[$S^1$:] not to send an assertion but to provide feedback, if possible,
\item[$S^2$:] to send an assertion but not to provide feedback, and
\item[$S^3$:] to send an assertion and to provide feedback, if possible.
\end{enumerate}

The feedback can be sent only if an assertion has been received. For this reason some elements in the matrix $M$ are identical:
\begin{enumerate}
\item $M_{00}=M_{10}=M_{01}=M_{11}$ (if no assertion has been transmitted, then no feedback can be sent), and, similarly,
\item $M_{02}=M_{03}$, $M_{12}=M_{13}$, $M_{20}=M_{30}$, $M_{21}=M_{31}$.
\end{enumerate}

We use $\{S_i^*,S_j^*\}$ to denote the Nash equilibrium strategy profile of the game. At each such equilibrium, we can find each actor's state variables such as their utilities, and normalized values of popularity, reputation, and knowledge. In general, despite a very special structure of $M$, the equilibrium in this game corresponds to a mixed strategy.

\section{\label{numerical}CASE STUDY: INFORMATION DISSEMINATION IN NETWORK}

In~\cite{zinoviev2010},  we explored the information dissemination process in a network with a large number of actors, based on the one-way two-player game model (of two interacting actors, only one was allowed to send an assumption, and the other could send feedback). We conducted a number of experiments by simulating the information dissemination in discrete time steps on a complete bidirectional social network of $1000$ actors. At each time step, one actor $A_i$ was randomly chosen to play the game with another randomly chosen actor $A_j$. The state values of both actors at Nash equilibrium for each game were recorded.

The model parameters of the simulations were selected based on the authors' common sense and were as follows. The probability of an assertion being true was $\varphi=0.8$. The actor popularity decay factor was $\delta=0.1$. The rumor discount coefficient was $\lambda=0.5$. The maximum number of assertions in the network was $N=2000$.

In each experiment, the network was populated by actors of a certain type. In the first experiment, all actors were ``Internet trolls'': $\kappa=0.1$, $\rho=0.1$, $\pi= 0.8$; in the second experiment, we modeled an ``Internet expert'' community: $\kappa=0.2$, $\rho=0.7$, $\pi= 0.1$. In each experiment, a third of the actors initially started as ``Ignoramuses'' ($k=0.1$), another third as ``Mediocres'' ($k=0.5$), and the remaining third as ``Gurus'' ($k=0.9$). The initial values of reputation $r$, popularity $p$, and the fractions of true and false assertions $f^+, f^-$ were drawn uniformly at random between the minimum value of $0$ and the maximum values of $0.5$ for $f^-$ and $1$ for the other three parameters.

\begin{figure}[tb!]
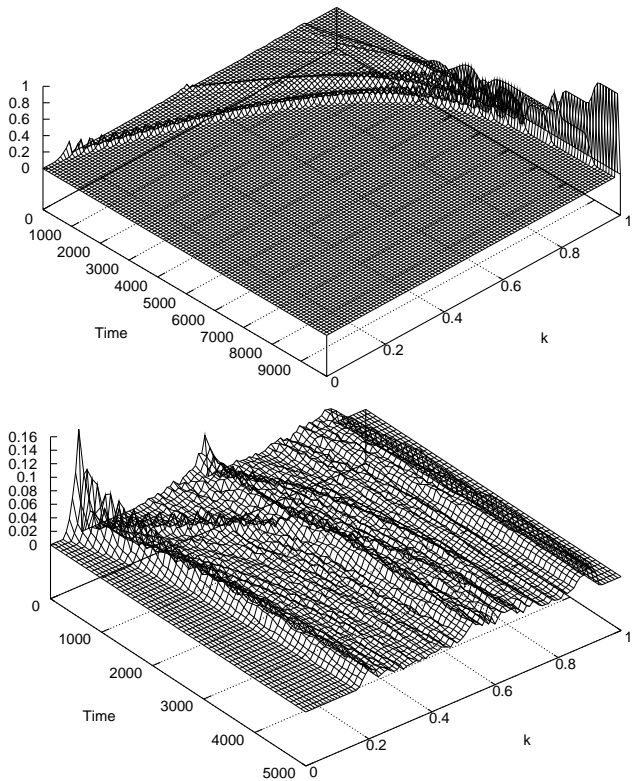
\centering
\epsfig{file=knowledge-trolls.epsi,width=\columnwidth}
\vskip0.5\baselineskip
\epsfig{file=knowledge-experts.epsi,width=\columnwidth}
\vskip0.5\baselineskip
\caption{\label{knowledge-troll}Distribution of actors ($z$ axis) by $k$ as a function of simulation time (one-way communications). Top: ``troll'' community; bottom: ``expert'' community. }
\end{figure}

For the duration of 10 million assertion transmissions (on average 10,000 communications per actor), we monitored the distribution of $k$ in the network as a function of the simulation time (the simulation time is defined as the average number of communications per actor). The results of  sample simulation runs are shown in Figure~\ref{knowledge-troll}. The most characteristic features of the graphs are the difference in learning speed between ``trolls'' (above) and ``experts'' (below) and the bifurcations in the ``expert'' community, where actors with higher reputation learn much slower than those with lower reputation.

\begin{figure}[tb!]
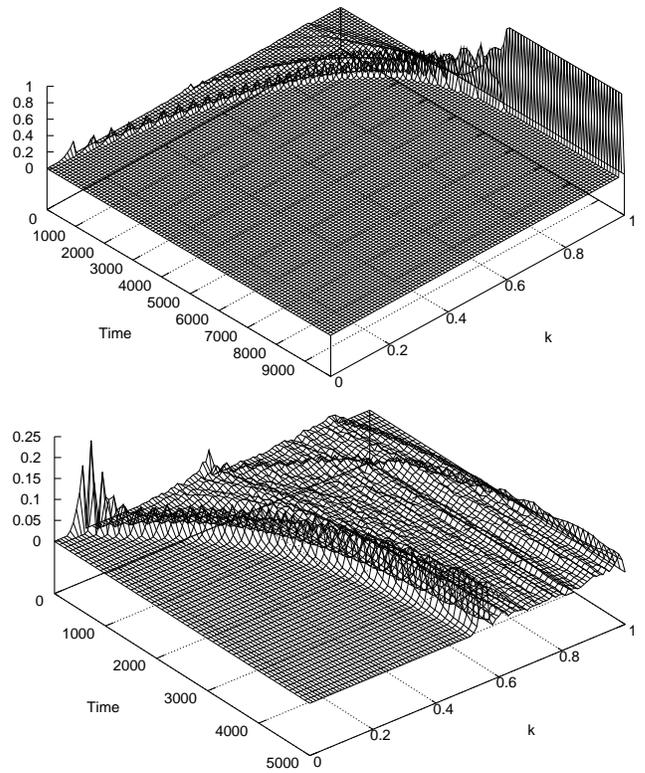
\centering
\epsfig{file=knowledge-trolls-duplex.epsi,width=\columnwidth}
\vskip0.5\baselineskip
\epsfig{file=knowledge-experts-duplex.epsi,width=\columnwidth}
\vskip0.5\baselineskip
\caption{\label{knowledge-troll-duplex}Distribution of actors  ($z$ axis) by $k$ as a function of simulation time (full-duplex communications). Top: ``troll'' community; bottom: ``expert'' community.}
\end{figure}

To understand the difference between one-way and two-way full-duplex communication models, we reproduced the original experiment  based on the four-strategy game model built in the previous section (Fig.~\ref{knowledge-troll-duplex}).

One can tell from Figs.~\ref{knowledge-troll} and \ref{knowledge-troll-duplex} that both models converge to stable states; however, the convergence speed and behavior are not the same. The ``troll'' communities converge to the state of ``total knowledge,'' where all actors become fully knowledgeable ($f\approx k\approx 1$), after a finite number of iterations. The ``full-duplex'' community converges about twice as fast as the ``one-way'' community because of the expanded capacity of the communication channels.

The distribution of information in the ``expert'' communities disperses slower in time then in the ``troll'' communities, but again faster when simulated using the two-way model. Most importantly, there is no more bifurcation for ``Ignoramuses'' and ``Gurus'' caused by the variation in their reputation. On the contrary, the low-reputation ``Mediocres'' soon become indistinguishable from ``Gurus,'' and high-reputation ``Mediocres'' join the ranks of the former ``Ignoramuses.''

\section{\label{emergence}CASE STUDY: EMERGENCE OF A SOCIAL NETWORK}

In another case study, we explored the possibility of emergence of social networks as the result of local optimization of the network members' utilities.

Our assumption is that a member $A_i$ of a social network $A$ engages in information exchange with her neighbors (``friends''). $A_i$ keeps the ``friendship'' relationship with $A_j$ if a communication with $A_j$ increases $A_i$'s utility. Otherwise, the relationship is severed after some time. We propose that the network built according to these principles is a ``small-world'' network (all major massive online social networks belong to this class).

\begin{figure}[tb!]\centering
\epsfig{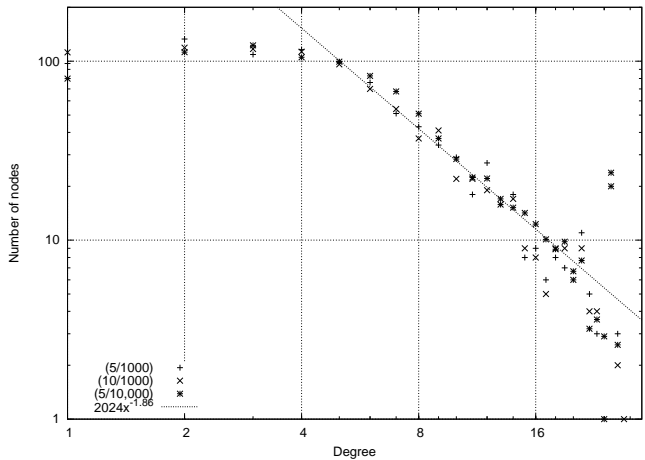}
\vskip0.5\baselineskip
\caption{\label{pareto}Distribution of node degree in an emerging social network.}
\end{figure}

To validate the hypothesis, we created a collection of 10,000 ``expert'' members. Then, for each member, we created 25 random ``friendship'' connections per member. We simulated five two-way interactions along each connection and accumulated the overall utility change for each connection. At the end, we discarded all connections with a negative utility. This process was repeated several times, but only the first iteration made a significant change to the network: the degree distribution in the range from 5 to 24 after the first iteration was close to Pareto distribution with the coefficient of $-1.86$, which is typical for a major real social network (Fig.~\ref{pareto}). Since Pareto distribution of node degrees is a common attribute of a small-world network, we treat this result as an implicit proof of our hypothesis.
\section{\label{conclusion}CONCLUSION AND FUTURE WORK}

In this paper we extended a previously developed game theoretical model of information dissemination. The model takes into account both personal traits of actors (the desire for self-perceived knowledge, reputation, and popularity) and the properties of the disseminated information (in particular, its overall truthfulness). The feedback mechanism is used to control the reputation of information senders and deter them from distributing unconfirmed rumors.

The model is mathematically represented by a two-player non-zero sum, non-cooperative game, where  the available actions are to forward an assertion or hold it indefinitely and to provide feedback on received assertions or not.

Our numerical experiments show that a massive social network of actors communicating using the proposed model demonstrates intuitively acceptable aggregate learning behavior, and that it is possible to explain an emergence of a ``small-world'' network by allowing its members to discard negative-utility connections.

To improve and generalize our model, we propose the following future research directions. 
\begin{itemize}
\item The variability of  $\kappa$, $\rho$, and $\pi$  for different actors in a network will be considered. 
\item A multicast scenario (when an actor atomically transmits an assertion to a group of recipients) will be analyzed.
\item A connection between $P$, $R$, and $K$ and selected characteristics of real massive online social networks will be explored.
\end{itemize}

\section*{Acknowledgment}
The authors are grateful to Honggang Zhang and Ali Yakamercan for their help with model building and writing simulation code. This research has been supported in part by the College of Arts and Sciences, Suffolk University, through an undergraduate research assistantship grant.

\bibliographystyle{acm}
\bibliography{cs}

\begin{thebibliography}{10}

\bibitem{NBERw14040}
{\sc Acemoglu, D., Dahleh, M.~A., Lobel, I., and Ozdaglar, A.}
\newblock Bayesian learning in social networks.
\newblock Working Paper 14040, National Bureau of Economic Research, May 2008.

\bibitem{crandall2008}
{\sc Crandall, D., Cosley, D., Huttenlocher, D., Kleinberg, J., and Suri, S.}
\newblock Feedback effects between similarity and social influence in online
  communities.
\newblock In {\em Proc. KDD'08\/} (Las Vegas, NV, Aug. 2008), ACM,
  pp.~160--168.

\bibitem{ellwardt2009}
{\sc Ellwardt, L., and {van Duijn}, M.}
\newblock Me and you and everyone we gossip about. a new direction in social
  networks.
\newblock Poster presentation at INSNA SunBelt, San Diego, CA, Mar. 2009.

\bibitem{gale2003}
{\sc Gale, D., and Kariv, S.}
\newblock Bayesian learning in social networks.
\newblock {\em Games and Economic Behavior 45}, 2 (2003), 329--346.

\bibitem{hirshman2009}
{\sc Hirshman, B., and Carley, K.}
\newblock Effect of cellular network structure on information diffusion.
\newblock Poster presentation at INSNA SunBelt, San Diego, CA, Mar. 2009.

\bibitem{1099206}
{\sc Lampe, C., and Johnston, E.}
\newblock Follow the (slash) dot: effects of feedback on new members in an
  online community.
\newblock In {\em GROUP '05: Proceedings of the 2005 international ACM SIGGROUP
  conference on Supporting group work\/} (New York, NY, USA, 2005), ACM,
  pp.~11--20.

\bibitem{nekovee2008}
{\sc Nekovee, M., Moreno, Y., Bianconi, G., and Marsili, M.}
\newblock Theory of rumour spreading in complex social networks.
\newblock {\em Physica A: Statistical Mechanics and its Applications 374}, 1
  (January 2007), 457--470.

\bibitem{parkhurst98}
{\sc Parkhurst, J., and Hopmeyer, A.}
\newblock Sociometric popularity and peer-perceived popularity.
\newblock {\em The Journal of Early Adolescence 18}, 2 (1998), 125--144.

\bibitem{spertus2005}
{\sc Spertus, E., Sahami, M., and Buyukkokten, O.}
\newblock Evaluating similarity measures: a large-scale study in the {O}rkut
  social network.
\newblock In {\em Proc. KDD'05\/} (Chicago, IL, Aug. 2005), ACM, pp.~678--684.

\bibitem{watts2002}
{\sc Watts, D., Dodds, P., and Newman, M.}
\newblock Identity and search in social networks.
\newblock {\em Science 296\/} (May 2002), 1302--1305.

\bibitem{webster1913}
{\sc Webster, N.}
\newblock {\em Webster's Revised Unabridged Dictionary}.
\newblock Merriam-Webster, Springfield, MA, 1913.

\bibitem{zanette2002}
{\sc Zanette, D.}
\newblock Dynamics of rumor propagation on small-world networks.
\newblock {\em Phys. Rev. E. 65\/} (2002), 041908.

\bibitem{zinoviev2010}
{\sc Zinoviev, D., Duong, V., and Zhang, H.}
\newblock A game theoretical approach to modeling information dissemination in
  social networks.
\newblock In {\em Proc. IMCIC 2010\/} (Florida, Apr. 2010), vol.~I, IIIS,
  pp.~407--412.

\end{thebibliography}
\end{document}